\documentclass[preprint,12pt,a4paper]{elsarticle}
\pdfoutput=1

\usepackage{graphicx}
\usepackage[usenames,dvipsnames]{xcolor}
\usepackage{fontenc,layout}
\usepackage[colorlinks=true]{hyperref}
\usepackage{epstopdf}
\usepackage{slashed}
\usepackage{subfig}
\usepackage {amsfonts}
\usepackage{amssymb}
\usepackage{amsmath}
\usepackage{comment}

 \DeclareMathOperator{\erf}{erf}
 
 \newcommand{\EWbasin}{\Delta_{EW}} % EW vacuum basin of attraction
 \newcommand{\UVbasin}{\Delta_{UV}} % UV vacuum basin of attraction
 \newcommand{\minEW}{h_{EW}} % EW vacuum
 \newcommand{\minUV}{h_{UV}} % UV vacuum
 \newcommand{\localmax}{h_{max}} % UV vacuum

\begin{document} 

\title{Higgs domain walls in the thermal background}

\author[a]{Tomasz~Krajewski}
\ead{Tomasz.Krajewski@fuw.edu.pl}
\author[a]{Zygmunt~Lalak}
\ead{Zygmunt.Lalak@fuw.edu.pl}
\author[b]{Marek~Lewicki}
\ead{Marek.Lewicki@adelaide.edu.au}
\author[a]{Pawe\l~Olszewski}
\ead{Pawel.Olszewski@fuw.edu.pl}
\address[a]{Institute of Theoretical Physics, Faculty of Physics, University of Warsaw,\\
ul. Pasteura 5, Warsaw, Poland}
\address[b]{Department of Physics, King's College London, Strand, WC2R 2LS London, UK}

\begin{abstract}
%Majority of cosmological models predict that the Universe in the past was hot i.e. filled with many particles (especially those from the Standard Model) with high kinetic energies. 
Most cosmological models predict that the universe was hot and dense at the early stages of it's evolution.
 In this paper we analyse the influence of the thermal bath of Standard Model particles on the dynamics of cosmological Higgs domain walls. This manuscript poses an~extension of our earlier work in which we investigated the evolution of networks of Higgs domain walls %in the background of the vacuum state.
neglecting the impact of temperature variation.

Using the thermally corrected effective potential of Standard Model we have found that both the position of the local maximum $h_{max}$ separating minima and the width of domain walls strongly depend on temperature $T$. For temperatures higher than $10^{10}\; \textrm{GeV}$ they respectively increase proportionally and decrease inverse proportionally to the increasing temperature. Thus, the energy scale of the problem follows the value of temperature. 

Our numerical lattice simulations based on the PRS algorithm reveal that Higgs domain walls in the presence of the background thermal bath are highly unstable and decay shortly after formation. Moreover we have found that the fraction of horizons produced by inflation in which Higgs field expectation value is higher then $h_{max}$ needs to be very low in order for the evolution of the~network of the domain walls to end in the electroweak vacuum. This means that Higgs domain walls necessarily were very rare objects and their average energy density was very small. As a result, the domain walls can not significantly effect cosmological observables. 
\\
\\
\\
\\
\\
KCL-PH-TH/2019-16
\end{abstract}

\begin{keyword}
domain walls \sep Higgs effective potential \sep reheating \sep inflation
\end{keyword}

\maketitle

\section{Introduction}
The measurement of the Higgs boson mass at the Large Hadron Collider \cite{Aad:2012tfa,Chatrchyan:2012ufa} started the new era of research of the cosmological evolution of the Higgs field. The observation of the existence of the second deeper minimum of the effective potential of the Standard Model (SM) for central measured values of its parameters took under consideration the question of the stability of the observed vacuum state of the Higgs field. Even through, it was determined \cite{Sher:1988mj,Casas:1994qy,Casas:1996aq,Casas:2000mn,Espinosa:2007qp,Ellis:2009tp,Degrassi:2012ry,Buttazzo:2013uya,Andreassen:2014gha,Lalak:2014qua,Salvio:2016mvj,Andreassen:2017rzq,Chigusa:2017dux,Markkanen:2018pdo} that the electroweak breaking minimum $\minEW$ is metastable in the flat Minkowski background, implications for cosmological history of the Higgs field expectation value are not yet completely understood. The evolution of the Higgs field during hypothetical inflationary epoch was widely discussed in the literature \cite{Espinosa:2007qp,Hook:2014uia,Espinosa:2015qea,East:2016anr,Darme:2017wvu,Rodriguez-Roman:2018swn,Markkanen:2018pdo,Franciolini:2018ebs}. % It was known for years \cite{Lalak:2007rs}%i pozostałe referencje
%that
Inflation produces an~approximately Gaussian distribution of the expectation value of light fields~\cite{Lalak:2007rs}, with the standard deviation $\sigma_I$ of the order of:  
\begin{equation}
\sigma_I \sim \frac{\sqrt{\mathcal{N}} H_I}{2 \pi},
\end{equation}
where $\mathcal{N}$ is number of e-folds and $H_I$ is the value of the Hubble parameter. In the case of the Higgs field, deviation from the Gaussian distribution is found for very large field strengths %(larger than the position of the local maximum $\localmax$ of the effective potential which separates the minima) 
due to existence of the second high field strength minimum $\minUV$. The existence (after inflation) of horizons with the Higgs field expectation value belonging to the basin of the attraction $\UVbasin$ of the high field strength minimum in our past light cone may be problematic from the point of view of cosmology~\cite{East:2016anr,Espinosa:2007qp}.% (fate of such horizons is still not fully understood).
In fact such horizons prove to be disastrous to our Universe, if they expand after inflation. Consequently, bounds on the value of the Hubble parameter $H_I$ during inflation have been proposed in order to avoid the field strengths beyond the potential barrier separating the two minima~\cite{Hook:2014uia}. On the other hand, the Higgs field may be stabilised during inflation either by the coupling to the inflaton field or the non-minimal coupling t gravity \cite{Hook:2014uia,Kamada:2014ufa,Espinosa:2015qea,Espinosa:2016nld,Markkanen:2018pdo}. In both cases the additional coupling produces, during inflation, an effective mass term for the Higgs field which suppresses high amplitude fluctuations. Even though, those two couplings stabilise the Higgs field during inflationary period, the problem will reappear after inflation, during (p)reheating period, when the effective mass term generated by the couplings decreases. %Unfortunately, the dynamics of the Higgs field during (p)reheating is poorly known due to model-dependence of the (p)reheating process.
%Finally, it was noticed \cite{Espinosa:2007qp,Kamada:2014ufa,Espinosa:2015qea,Espinosa:2016nld,Franciolini:2018ebs} that background of thermal bath of the Standard Model particles stabilise the Higgs field due to enlarging of the basin of attraction $\EWbasin$ of the electroweak vacuum (the position of the local maximum $\localmax$ is shifted towards larger values) in the reheated Universe. We discuss this effect in details in section \ref{SM_potential}.
However, the background of thermal bath of the Standard Model particles also has a positive effect as it stabilises the Higgs field~\cite{Espinosa:2007qp,Kamada:2014ufa,Espinosa:2015qea,Espinosa:2016nld,Franciolini:2018ebs} due to enlarging of the basin of attraction $\EWbasin$ of the electroweak vacuum (the position of the local maximum $\localmax$ is shifted towards larger values) in the reheated Universe. We discuss this effect in details in section \ref{SM_potential}.

In our previous papers \cite{Krajewski:2016vbr,Krajewski:2017czs} we investigated the evolution of the Higgs field in the radiation domination epoch assuming that the Universe cooled down (before the time under consideration) to the temperature low enough for thermal corrections to be negligible. We examined formation and decay of domain walls existing in configuration of the Higgs field. In this paper we are interested in the situation when the evolution of Higgs domain walls takes place in the background of the thermal bath of SM particles with high kinetic energies. We will show that properties of Higgs domain walls in high temperatures differs from previously computed zero-temperature case. We perform a series of numerical simulations of the evolution of Higgs domain walls with various initial values of the temperature. Our results show that Higgs domain walls in the thermal background with high temperature are highly unstable and their production in significant amounts in the early Universe is excluded by the requirement that the final state of their evolution has to be the electroweak vacuum. In contrast to previous papers in which we were not interested in how the initial inhomogeneous configuration of the Higgs field was formed during the evolution of the Universe, in this paper we will assume the inflationary scenario and will relate our parameters such as the reheating temperature $T_{RH}$ and the value of the Hubble parameter during inflation $H_I$. Even though, some of our results do not depend on this assumption and are valid independently on the specific model of the early Universe evolution.

The most important questions which we address in this paper are if the formation of Higgs domain walls in the early Universe reheated to high temperatures is consistent with experimental data and if taking into account their dynamics can help in weakening of the simplified bound on the Hubble parameter $H_I$ during inflation. Our main result is the observation that the domain walls in the thermal background are highly unstable. In consequence only a very small amount of them is allowed in the early Universe. Thus, the simplified bound on $H_I$ resulting from the assumption that none of them were formed is not modified significantly.

The paper is organised as follows. In section \ref{SM_potential} we briefly present the form of the thermally corrected effective potential of the SM and discuss the influence of the thermal corrections on its shape. %We summarize present knowledge on the role of the thermally induced fluctuations of the Higgs field in section \ref{initialization}. 
In section \ref{wall} we show how properties of Higgs domain walls are modified by thermal corrections.The determination of the space of initial parameters for which the Higgs domain walls might be formed in the early Universe is performed in section \ref{initial}. Section~\ref{decay_time} presents result of our numerical simulations. Our results are summarized in section \ref{summary}.

\section{Thermal corrections to SM effective potential\label{SM_potential}}
In our two previous papers \cite{Krajewski:2016vbr,Krajewski:2017czs} we investigated the evolution of domain walls of the Higgs field in the SM and its extensions. In our attempt to understand the role of the networks of Higgs domain walls in the history of the Universe we have been working under simplifying assumption that SM particles present in early Universe had negligible effect on the dynamics of the expectation value of the Higgs field. This manuscript is dedicated to study of opposite regime, specifically the evolution of the Higgs field in the background of the thermal bath of interacting particle. Extrapolating back in time our present knowledge on the Universe we deduce that it was much denser and hotter i.e. filled with many particles with high kinetic energies, in the past. %The thermal background state with high temperature $T$ is more adequate for calculations in this situation than the vacuum state used previously.
As shown by many authors \cite{Espinosa:2007qp,Kamada:2014ufa,Espinosa:2015qea,Espinosa:2016nld,Franciolini:2018ebs} the dynamics of the Higgs field in the thermal state differs significantly from the case of the vacuum for large enough temperatures.

Various formalisms for Thermal Quantum Field Theory have been developed in order to investigate different aspects of thermal processes. Assuming that the SM particles present in the early Universe stayed in the thermal equilibrium during the evolution of Higgs domain walls we have used the Matsubara's imaginary time formalism for computing the thermally corrected effective potential of the SM $V_{\text{eff}} (h, T)$. Our calculations based directly on the methods presented in \cite{Delaunay:2007wb,Lewicki:2016efe}. In this study thermal corrections to the effective potential were included at the 1-loop order using the formalism of external field in the form:
\begin{equation}
\begin{split}
V_{\text{SM}} (h, T) &= \frac{1}{2} m^2 h^2 + \frac{1}{4} \lambda h^4 + \sum_{i=\varphi,\chi,W,Z,t} n_i \frac{{m_i}^4 (h)}{64 \pi^2} \left[\log \frac{{m_i}^2(h)}{\mu^2} - C_i\right]\\
&+ \sum_{i=\varphi,\chi,W,Z} \frac{n_i T^4}{2 \pi^2} J_b \left(\frac{{m_i}^2(h)}{T^2}\right) + \frac{n_t T^4}{2 \pi^2} J_f \left(\frac{{m_t}^2(h)}{T^2}\right)\\
&+ \sum_{i=\varphi,\chi,W,Z,\gamma} \frac{\bar{n}_i T}{12 \pi} \left[{m_i}^3(h) - \left({m_i}^2(h) + \Pi_i (T) \right)^{\frac{3}{2}} \right],
\end{split} \label{eq:thermal_potential}
\end{equation}
where $m_i$ are field-dependent masses:
\begin{subequations}\label{higgs_masses}
\begin{align}
{m_h}^2 (\phi) &= m^2 + 3 \lambda h^2,\\
{m_{\chi_i}}^2 (\phi) &= m^2 + \lambda h^2,\\
{m_W}^2 (\phi) &= \frac{g^2}{4} hi^2,\\
{m_Z}^2 (\phi) &= \frac{g^2 + g'^2}{4} h^2,\\
{m_t}^2 (\phi) &= \frac{{y_t}^2}{2} h^2.
\end{align}
\end{subequations}
$\Pi_i (T)$ are Debye's masses resulting from \emph{ring-improvement} of the finite temperature potential (see \cite{Delaunay:2007wb,Lewicki:2016efe} for details) and $\varphi,\chi,W,Z,t$ are respectively Higgs field fluctuations in the direction parallel to the expectation value $h$ and orthogonal to it, $W$ and $Z$ bosons and the top quark. We have used following tree-level values of the coefficients: $n_{\{\varphi, \chi, W, Z, t\}} = \{1,3,6,3,-12\}$, $\bar{n}_{\{\varphi, \chi, W, Z, t\}} = \{1,3,2,1,1\}$ and $C_{\varphi, \chi, t}=\frac{3}{2}$, $C_{W, Z}=\frac{5}{6}$. Finally, functions $J_{b/f}$ are given in the integral form:
\begin{equation}
J_{b/f} (x):= \int_0^\infty dk k^2 \log{\left[ 1 \mp e^{-\sqrt{k^2 +x}} \right]}.\label{J_definition}
\end{equation}
For our numerical studies we have used implementation of function $J_{b/f}$ documented in \cite{Fowlie:2018eiu}.

\begin{figure}[!ht]
\centering
\includegraphics[width= 0.7\textwidth]{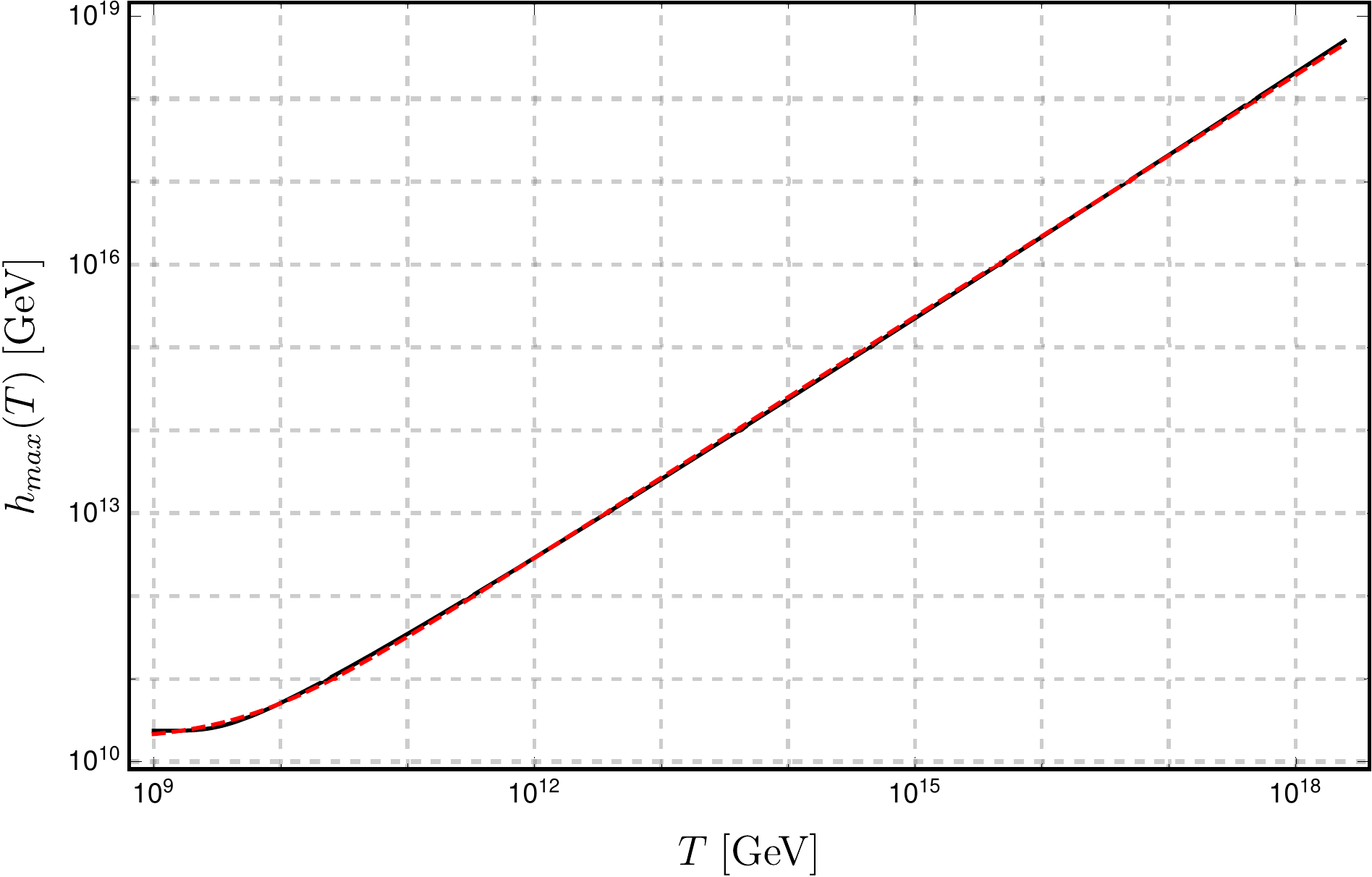}
\caption{Position of the local maximum $\localmax (T)$ of the thermally corrected effective potential $V_{\text{eff}} (h, T)$ as a function of the temperature $T$.\label{thermal_maximum_plot}}
\end{figure}

One of important parameters in investigating the dynamics of networks of Higgs domain walls %(among others used for detection of domain walls in our numerical code) 
is the position of the local maximum $\localmax$ separating minima of the potential. The dependence of the field strength $\localmax (T)$ at which the thermally corrected effective potential of SM $V_{\text{eff}} (h, T)$ develops the local maximum on the value of the temperature $T$ is presented in the figure \ref{thermal_maximum_plot}. For temperatures lower than $10^9\; \textrm{GeV}$ thermal corrections do not influence $\localmax$ and its value is equal to the vacuum case. However, for higher temperatures $\localmax$ grows approximately linearly with the temperature $T$. Dashed line on the plot \ref{thermal_maximum_plot} correspond to the fitted function:
\begin{equation}
\localmax(T)= a~T^b + c, \label{analytic_max}
\end{equation}
with the best-fit parameters $a$, $b$ and $c$ presented in the table \ref{max_fit}. The value of the exponent $b$ is close to $1$ which confirms that the $\localmax (T)$ grows nearly linearly with the temperature $T$. Moreover, thermal corrections are negligible for temperatures $T \ll c$. We used this fitted function \eqref{analytic_max} rather than the full expression in our numerical simulations for the sake of performance.

\begin{table}
\centering
\begin{tabular}{|c|c|c|}\hline
\multicolumn{3}{|c|}{parameters}\\ \hline
$a$ & $b$ & $c$ \\
 \hline
 $6.08 \pm 0.21$ & $0.9723 \pm 0.0010$ & $(1.783 \pm 0.043)\times 10^{10} $ GeV \\ \hline
\end{tabular}
\caption{Parameters $a$, $b$ and $c$ for the model given by \eqref{analytic_max} fitted to the dependence of the position of local maximum $\localmax (T)$ on the value of the temperature $T$. \label{max_fit}}
\end{table}

The shift of the position $\localmax$ of the local maximum due to thermal corrections stabilizes the electroweak vacuum in the cosmological context. With increasing temperature the basin of attraction $\EWbasin$ of electroweak vacuum grows and field strength values that belongs to the basin of attraction $\UVbasin$ of high field strength minimum $\minUV$ at zero temperature lay in $\EWbasin$ for high enough temperature $T$. As a result, higher amplitudes of fluctuations of the Higgs field are allowed for higher temperatures. Thus, initial distributions of the Higgs field that was found in our previous studies to decay into excluded $\minUV$ when evolving in the background of the vacuum state may decay into $\minEW$ for high temperatures of the background state. 

\begin{figure}[!ht]
\begin{minipage}[t]{0.49\textwidth}
\includegraphics[width=\textwidth]{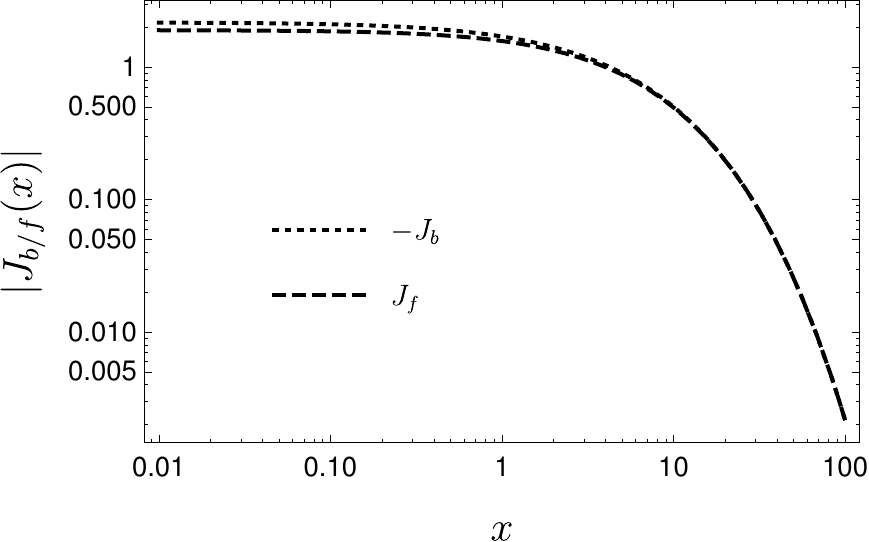}
\caption{Functions $J_f$ and $J_b$ defined in eq. \eqref{J_definition} present in the expression \eqref{eq:thermal_potential} for thermally corrected potential of the SM.\protect\label{J_functions}}
\end{minipage}
\hspace{0.5cm}
\begin{minipage}[t]{0.49\textwidth}
\includegraphics[width=\textwidth]{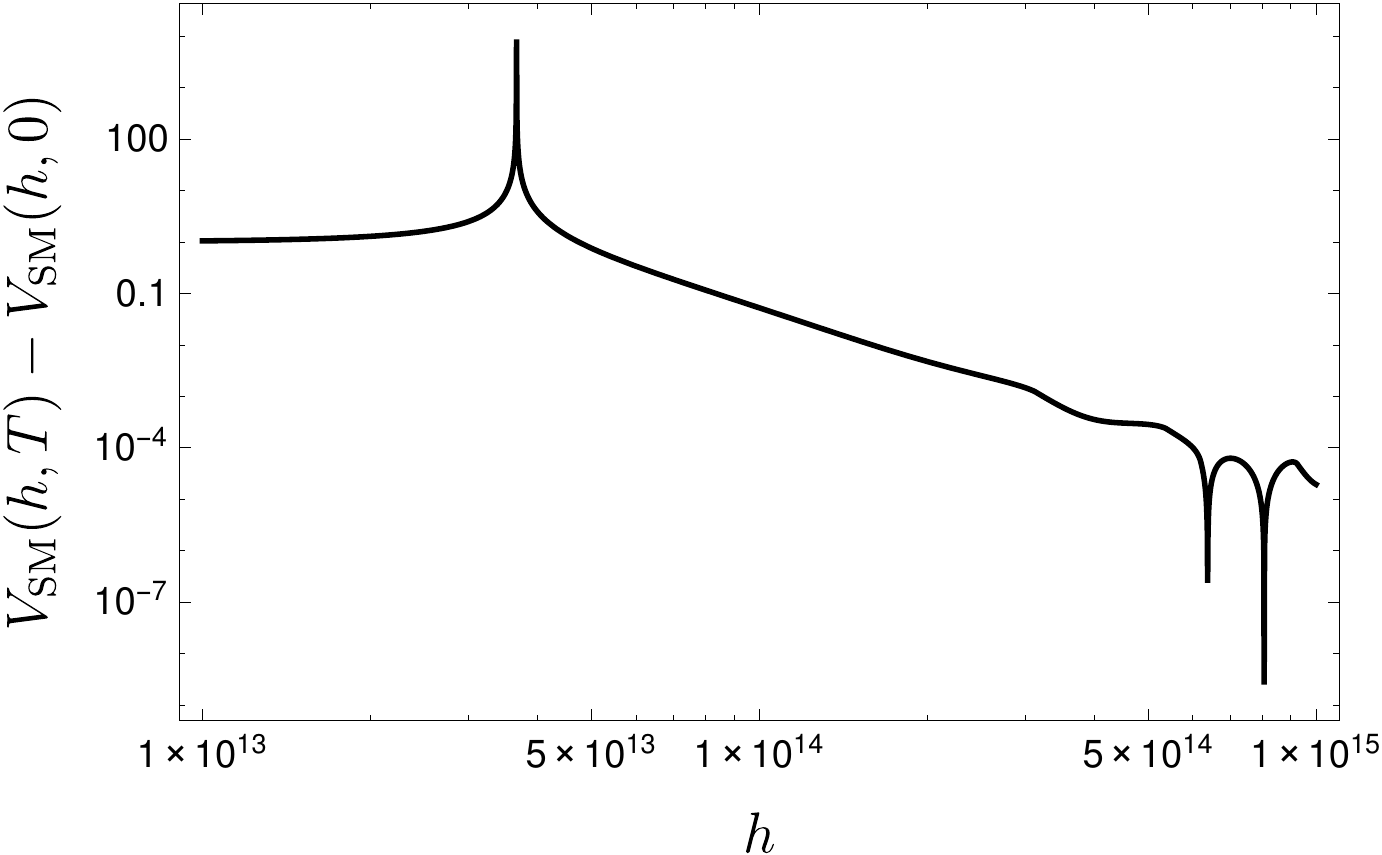}
\caption{Thermal corrections to the effective potential of the SM $\frac{V_{\text{SM}} (h, T) - V_{\text{SM}} (h, 0)}{V_{\text{SM}} (h, T)}$ for temperature equal to $T=10^{13}\; \textrm{GeV}$ as a~function of the Higgs field strength~$h$. \protect\label{thermal_corrections}}
\end{minipage}
\end{figure}

The second important quantity describing the shape of the potential $V_{\text{eff}} (h, T)$ is the difference of its values at its minima $V_{\text{eff}} (\minEW, T) - V_{\text{eff}} (\minUV, T)$. The functions $J_{b/f}$ (plotted in the figure \ref{J_functions}), become quickly very small with increasing of their argument. Arguments of functions $J_{b/f}$ in the expression \eqref{eq:thermal_potential} for thermally corrected potential of the SM are masses divided by the temperature $T$ of the thermal bath.  Thermally corrected masses (given by eq. \eqref{higgs_masses}) are increasing functions of the Higgs field strength so corrections are highly suppressed for values of the Higgs field strength higher than the value of the temperature $T$. As a~result, thermal corrections are negligible for very high field strengths and the value of effective potential $V_{\text{SM}} (\minUV, T)$ at the high field strength minimum $\minUV$ stays unchanged. The ratio of the value of thermal correction to the value of the effective potential $(V_{\text{SM}} (h, T) - V_{\text{SM}} (h, 0))/V_{\text{SM}} (h, T)$ as a~function of the Higgs field strength is shown in figure~\ref{thermal_corrections} for exemplary value $T=10^{13}\; \textrm{GeV}$. The presented plot shows that thermal corrections drop quickly for high field strengths and the value of thermally corrected effective potential $V_{\text{SM}} (h, T)$ is nearly equal its value at zero temperature $V_{\text{SM}} (h, 0)$.

The magnitude of thermal corrections to the derivative of the effective potential drops quickly with the increasing Higgs field strength $h$ and for high enough field strengths become smaller than the uncertainty of calculation of the value of the derivative $\frac{\partial V_{\text{SM}}}{\partial h} (h, T)$ by itself due to experimental uncertainties of the measurement of the top quark mass $M_t$ and Higgs boson mass $M_h$. We conservatively estimate this uncertainty as: 
\begin{multline}
\sigma_{\partial V} (h, T) := \max\left(\left|\frac{\partial V_{\text{SM}}^{M_t+\sigma_{M_t},M_h}}{\partial h}(h, T) - \frac{\partial V_{\text{SM}}^{M_t-\sigma_{M_t},M_h}}{\partial h}(h, T)\right|,\right.\\
\left.\left|\frac{\partial V_{\text{SM}}^{M_t,M_h+\sigma_{M_h}}}{\partial h}(h, T)-\frac{\partial V_{\text{SM}}^{M_t,M_h-\sigma_{M_h}}}{\partial h}(h, T)\right|\right), \label{error_estimate}
\end{multline}
where $\frac{\partial V_{\text{SM}}^{M_1,M_2}}{\partial h}(h, T)$ is the value of the derivative for the Higgs field strength $h$ and temperature $T$ calculated for the top quark mass equal to $M_1$ and the Higgs boson mass to $M_2$. 
We independently model the thermal corrections $\frac{\partial V_{\text{SM}}}{\partial h} (h, T) - \frac{\partial V_{\text{SM}}}{\partial h} (h, 0)$ to the derivative of the effective potential and the zero-temperature contribution $\frac{\partial V_{\text{SM}}}{\partial h} (h, 0)$. The later we model as was previously described in \cite{Krajewski:2016vbr}.  Due to increase in computational complexity in numerical modelling of function of two variables instead of one we had to use less accurate model for thermal corrections than for the zero-temperature derivative. We used the measure $\sigma_{num}$ of uncertainty of our numerical model $\frac{\partial V}{\partial h}_{num}$ of $\frac{\partial V_{\text{SM}}}{\partial h}$ defined as:
\begin{equation}
\sigma_{num}\left[\frac{\partial V}{\partial h}_{num}\right]=\sup_{
\begin{array}{l}
h\in [10^{-6} T_0, 2 M_{Pl}] \\
T \in [10^{3} T_0, 10 T_0]
\end{array}
}\frac{\left|\frac{\partial V}{\partial h}_{num}(h, T)-\frac{\partial V_{\text{SM}}}{\partial h}(h,T)\right|}{\sigma_{\partial V} (h, T)},
\end{equation}
where $T_0$ is the initial value of temperature in our simulation. Values of $\sigma_{num}$ for numerical models used in our simulations do not exceeded $50\%$.

\section{Properties of SM domain walls in the background of thermal bath\label{wall}}
Imaginary time formalism allows for computation of observables in the constant thermal state background. On the other hand we are interested in the dynamics of the Higgs field in the evolving Universe. Thus, we need to assume that the time scale of interactions of Standard Model particles forming the thermal background state is much shorter than the Hubble scale $H^{-1}$, so they stay in the thermal equilibrium through the cosmological evolution of the Universe at least for the period of the existence of Higgs domain walls. Under this assumption we can approximate the equation of motion for the time-space dependent expectation value of the Higgs field as follows:
\begin{equation}
\frac{\partial^2 \phi}{\partial \eta^2} + \frac{2}{a} \left(\frac{d a}{d \eta}\right) \frac{\partial \phi}{\partial \eta} - \Delta \phi + a^2 \frac{\partial V_{\text{SM}}}{\partial \phi}(\phi, T (\eta))=0, \label{SM_eom}
\end{equation}
where $\phi$ is a~real scalar field which models the Higgs field in our simulations and $V_{\text{SM}}$ is the thermally corrected effective potential of the SM in temperature $T$ and $\Delta \phi$ denotes the Laplacian of the field $\phi$. For the purpose of this paper we consider the gravitational background in the form of the Friedman-Robertson-Walker metric background:
\begin{equation}
g = dt^2 - a^2(t) \delta_{ij} dx^i dx^j = a^2(\eta) \left(d\eta^2 -\delta_{ij} dx^i dx^j\right), 
\end{equation}
where Latin indices correspond to spatial coordinates, $t$ is cosmic time and $\eta$ denotes conformal time (such that $d \eta = \frac{1}{a(t)} dt$), 

The most reliable %(and the most complex)
 method of investigating the dynamics of cosmological domain walls available so far is solving the eom of the symmetry breaking field in lattice simulations. %We have used it while studying the evolution of Higgs domain walls in the thermal background.
  We adopt it by solved eq. \eqref{SM_eom} in numerical simulations on the 3D lattice, based on the PRS algorithm \cite{Press:1989yh}. The knowledge of a~length (or equivalently energy)\footnote{We are using the convention $\hbar=1=c$, so lengths and times are expressed in units of the inverse of the energy ($\textrm{GeV}$).} scale of the problem is needed in order to properly set lattice simulations. In the case of networks of cosmological domain walls two length scales are important: the typical size of domain and the width of domain walls $w$. The former changes during the evolution of the network and its growth in late times limits the dynamical range of the simulation. On the other hand, the later stays constant in time. In \cite{Krajewski:2016vbr} we have proposed the algorithm for computation of the width of domain walls for a~generic potential. It is based on the concept of finding a~time independent and planar (i.e. translationally invariant in two directions) solution of the eom in the Minkowski background. Such solution can be found using the first integral method as described in details in \cite{Krajewski:2016vbr}. Then, we estimate the width of domain walls as the~distance from the centre of the solution approximating the domain walls (the point at which the field strength equals the position of the local maximum of the potential) in the direction perpendicular to the surface of the wall at which the majority of the potential energy is stored. In the past, simulations with the physical width of walls varying from 2 to 100 lattice spacing $l$ (i.e. the physical distance between neighbouring points) were performed \cite{Press:1989yh,Coulson:1995uq,Lalak:1996db,Oliveira:2004he,Lalak:2007rs,Kawasaki:2011vv,Leite:2011sc,Hiramatsu:2013qaa}.

\begin{figure}[t]
\centering
\includegraphics[width= 0.7\textwidth]{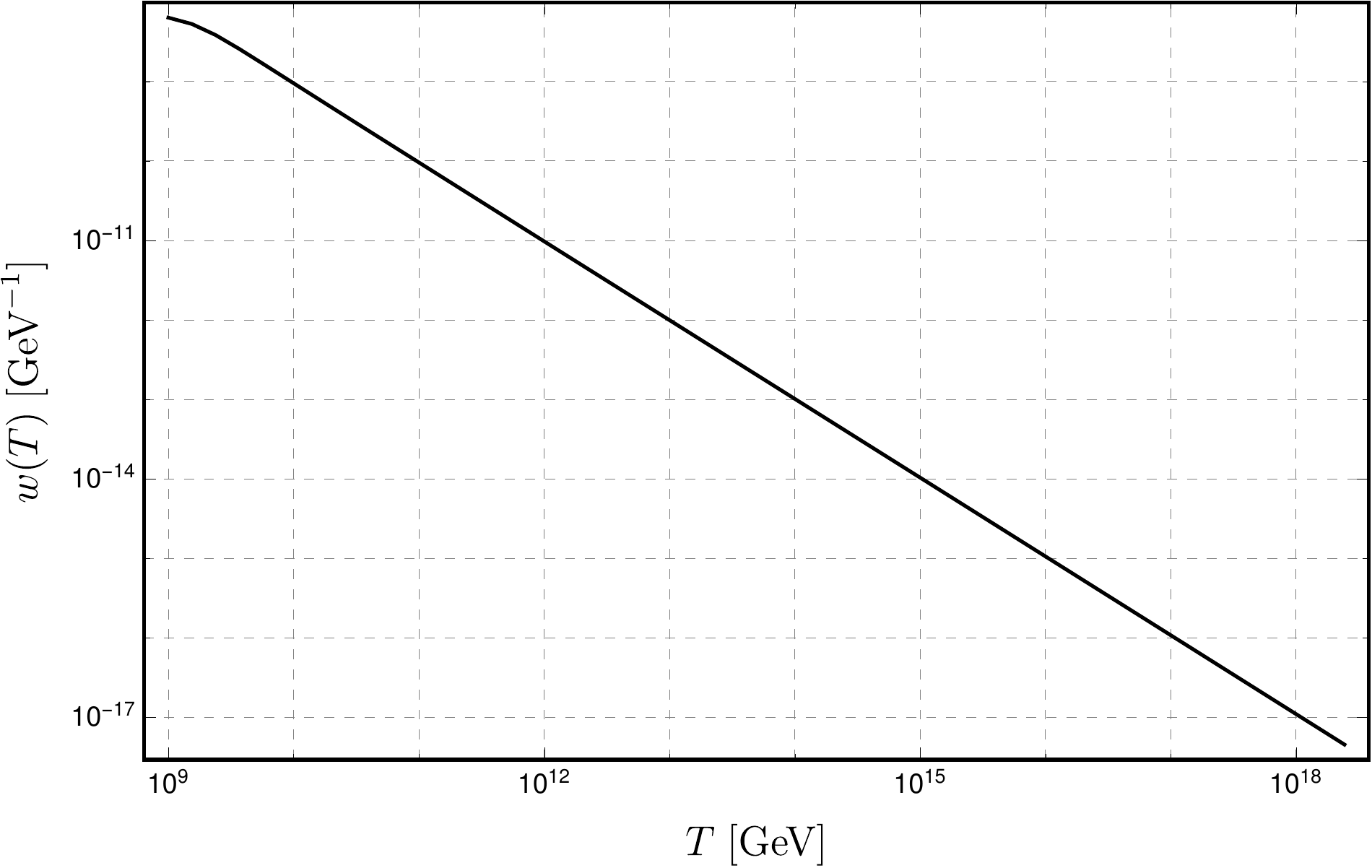}
\caption{The width of domain walls $w(T)$ as a~function of the temperature $T$ of the thermal bath.\label{width_thermal}}
\end{figure}

In the section \ref{SM_potential} we have shown that the position of the local maximum $\localmax$ of the thermally corrected effective action of the Standard Model does not change for low temperatures (lower than $10^9\; \textrm{GeV}$). On the other hand, for high temperatures $\localmax$ grows linearly with temperature $T$. Thus, it can be presumed that the characteristic scale of the dynamics of Higgs domain walls are not modified as far as temperatures are low enough and for high temperatures is of the order of $T$. This conjuncture finds its confirmation in the determination of the width of domain walls performed using described algorithm for the thermally corrected effective action of the SM \eqref{eq:thermal_potential}. The obtained dependence of the width $w (T)$ on temperature of the thermal bath $T$ is presented in figure \ref{width_thermal}. Calculated value of the width are nearly constant for temperatures lower than $10^9\; \textrm{GeV}$ and is close to the zero temperature result, obtained previously \cite{Krajewski:2016vbr}. For high temperatures the width drops approximately as $T^{-1}$.

%We have chosen the physical lattice spacing $l$ to be equal to $10^{-10}\; \textrm{GeV}^{-1}$ which leads to widths of walls contained in the range $35\, l \le w(\Lambda) \le 50\, l$. Henceforth, we will often use $\frac{1}{l} = 10^{10}\; \textrm{GeV}$ as a~unit of energy and inverse distance in space-time.

\section{Initial conditions for simulations\label{initial}}
Inflationary models predict that the Higgs field expectation value during inflation can be statistically described by the probability distribution $P(h)$ with evolution described by the Fokker-Planck equation. Solutions of the Fokker-Planck for the Higgs field \cite{Coulson:1995uq,Espinosa:2007qp,Hook:2014uia,Espinosa:2015qea,East:2016anr}, in the first approximation, are of the form of the Gauss distribution:
\begin{equation}
P(h) = \frac{1}{\sqrt{2 \pi} \sigma_I} e^{-\frac{\left(h -\theta\right)^2}{2 \sigma_I^2}}, \label{gauss_distribution}
\end{equation}
with the standard deviation of the order of
\begin{equation}
\sigma_I \sim \frac{\sqrt{\mathcal{N}} H_I}{2 \pi}, \label{inflationary}
\end{equation}
where $H_I$ is the value of the Hubble constant during inflation and $\mathcal{N}$ is the number of e-folds. The Hubble constant $H_I$ is the main parameter of inflationary models. Solving majority of problems found in the standard cosmology requires at least from $50$ to $60$ e-folds, so the standard deviation $\sigma_I$ can be determined in an~inflationary model. Unfortunately, evolution of $P(h)$ turns out to be independent of the mean value of the Higgs field and this value did not changed significantly during inflation. As a~result, the mean value of the Higgs field after inflation $\theta$ cannot be determined in an~inflationary model and a~priori it can take any value from $0$ to the Planck scale. As we argued in the section \ref{SM_potential} the position of the local maximum $\localmax$ grows approximately linearly with the temperature $T$ for $T>10^{11}\; \textrm{GeV}$, while the width of domain walls decreases as $T^{-1}$. Thus, we suppose that the the dynamics of Higgs domain walls for temperatures larger then $T=10^{9}\; \textrm{GeV}$ will differ from the previously considered vacuum case. Its energy scale should grows for high temperatures, so the evolution of networks of domain walls will be faster and start earlier. In our numerical simulation we investigate the range of initial temperatures $T_{start}$ from $10^{10}\; \textrm{GeV}$ to $10^{18}\; \textrm{GeV}$.

%%%%%%%%%%%%%%%%%%%%%%%%%%%%%%%%%%%%%%%%%%%%%%%%%%%%%%%%%%%%%%%%%
%%%%%%%%%%%%%%%%%%%%%%%%%%%%%%%%%%%%%%%%%%%%%%%%%%%%%%%%%%%%%%%%%
%%%%%%%%%%%%%%%%%%%%%%%%%%%%%%%%%%%%%%%%%%%%%%%%%%%%%%%%%%%%%%%%%

%Our simulations were initialised at the conformal time $\eta_{start}$ ranging from \mbox{$10^{-2}\, l = 10^{-12}\; \textrm{GeV}^{-1}$} to \mbox{$1 \, l = 10^{-10}\; \textrm{GeV}^{-1}$}. In our previous work~\cite{Krajewski:2016vbr} we showed that taking initial times below this interval does not modify the results. On the other hand, cosmological domain walls whose evolution we model in our numerical simulations, need to be superhorizon at the initialisation (i.e. width of domain walls must be larger than the Hubble horizon $H(\eta_{start})^{-1} \sim a(\eta_{start}) \eta_{start}$ at the time of initialisation $\eta_{start}$). The conformal time when domain walls are formed in the early Universe, must be greater than $\eta_{start}$ for the the initial numerical fluctuations to be smoothed out by the evolution of the field.
%The majority of cosmological models predict that the early Universe was (at least for a part of its history) hot and dense i.e. filled with many particles with high kinetic energy. Thus, it is reasonable to investigate the influence of the thermal background state on the dynamics of the Higgs domain walls. 

\begin{figure}[!ht]
\subfloat[]{\label{thermal_sigmamax}
\includegraphics[width=0.5 \textwidth]{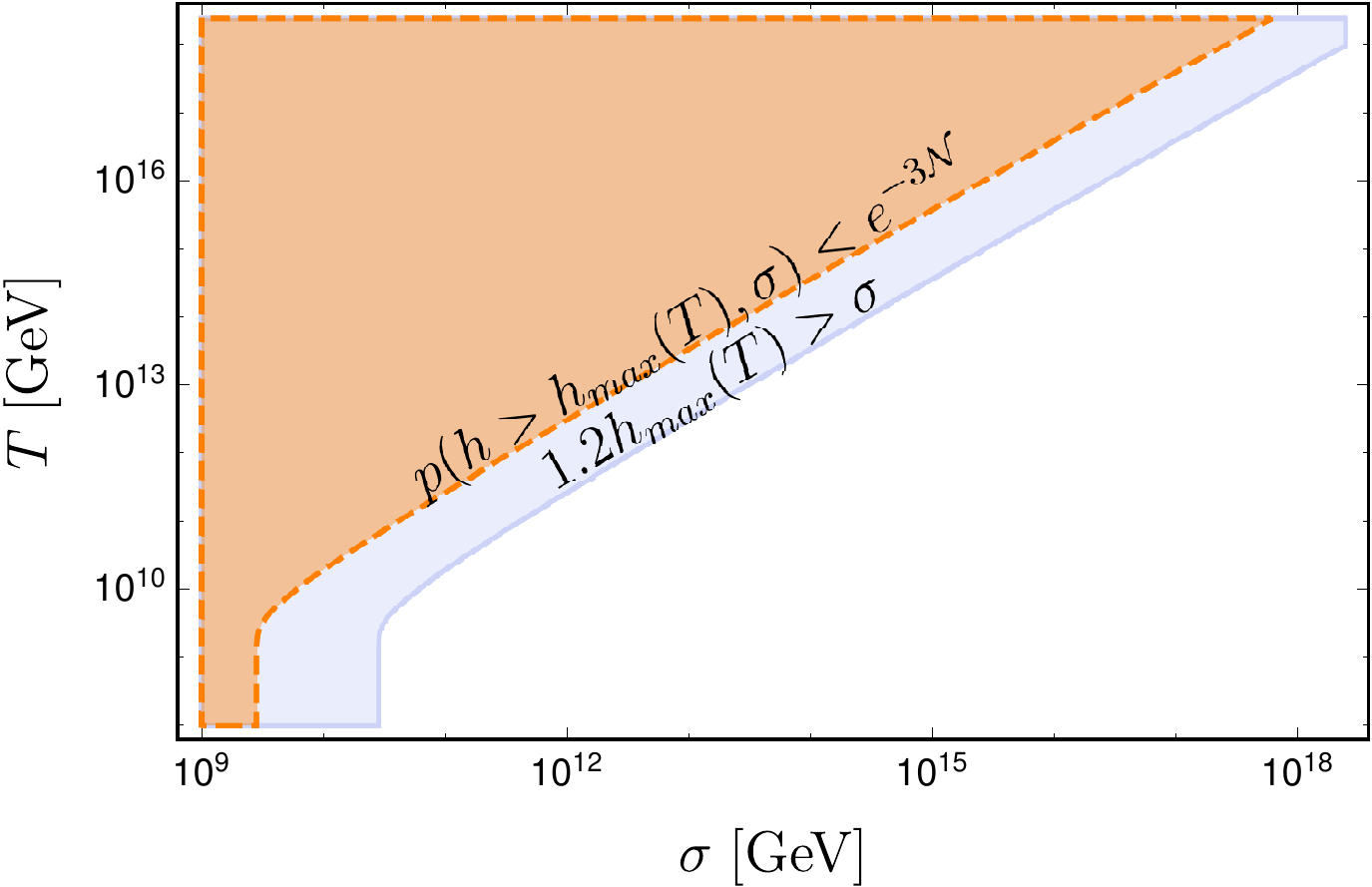}
 }
 \subfloat[]{\label{thermal_Hmax}
\includegraphics[width=0.5 \textwidth]{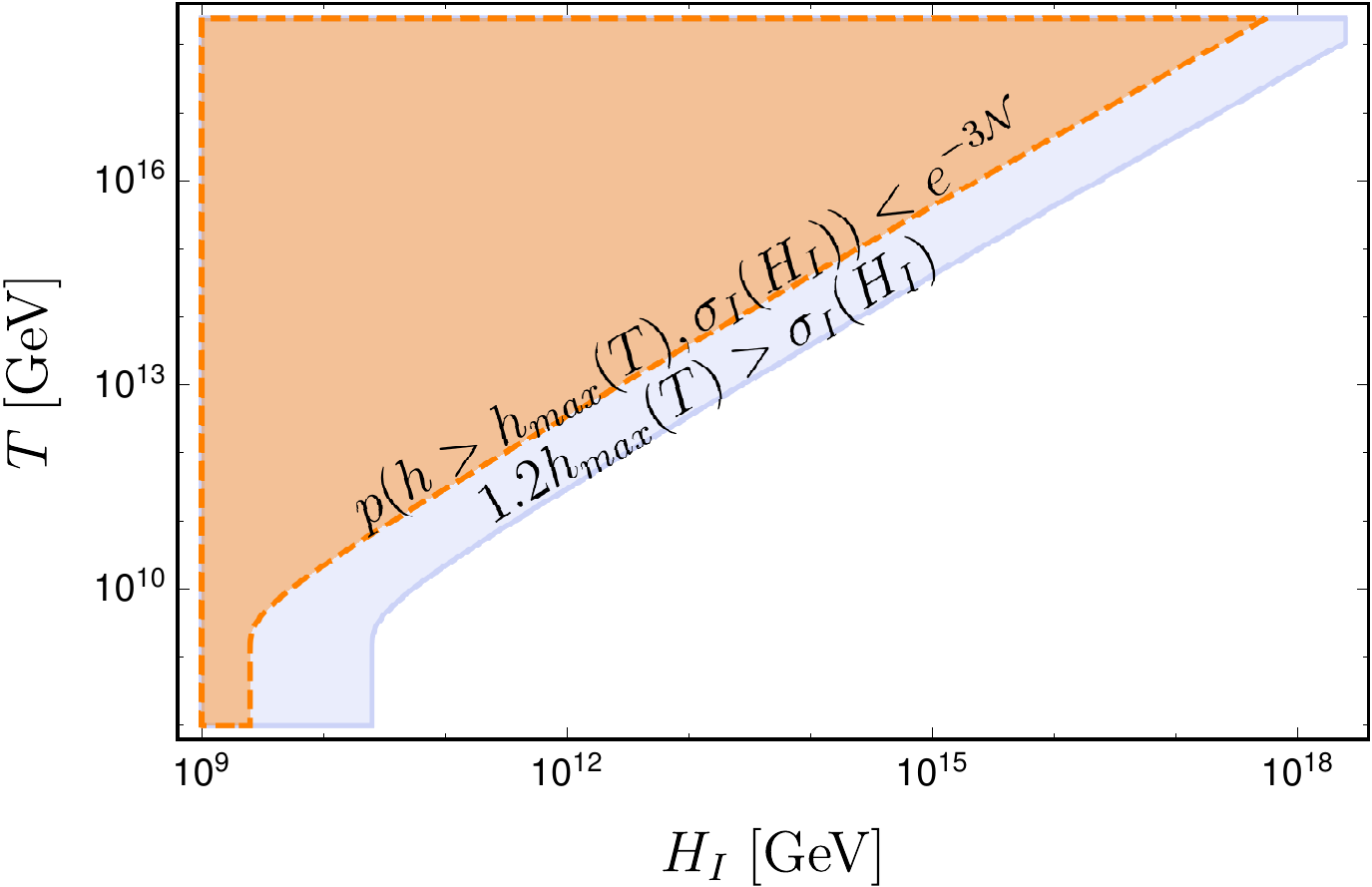}
 }
\caption{The range of the temperatures for which $1.2 \localmax (T) > \sigma$ (blue region) and \mbox{$1-\frac{1}{2}\erf\left(\frac{\localmax(T)}{\sqrt{2}\sigma}\right) < e^{-3\mathcal{N}}$} (orange region) \protect\subref{thermal_sigmamax}. Plot \ref{thermal_Hmax} presents these regions as a~function of the Hubble parameter $H_I$ under assumption: $\sigma_I =\frac{H_I}{2 \pi} \sqrt{\mathcal{N}}$. Domain walls can only develop with parameters corresponding to the blue region and not covered by the orange one.
 \protect\label{thermal_maximal}}
\end{figure}

Extrapolating previous results for the case of the zero temperature effective potential described in \cite{Krajewski:2016vbr,Krajewski:2017czs} we can expect that Higgs field configurations from probability distributions with $\sigma <1.2 \localmax$ and~$\theta=0$ should evolve toward the electroweak vacuum. Thus, we can estimate the range of temperatures for which end point of evolution of our network networks will be cosmologically acceptable. Moreover, the formation of domain walls can happen only when the fraction of the volume occupied by the field strength belonging to the basin of attraction of the electroweak vacuum $\EWbasin$ is of the order of $\mathcal{O}(1)$, so it should be $\sigma \sim \localmax$. The blue region on the plot in the figure \ref{thermal_sigmamax} corresponds to the bound $1.2 \localmax (T) > \sigma$. While the orange region corresponds to the range of temperatures for which the following condition is satisfied:
\begin{equation}
p(|h|>\localmax(T),\sigma) \approx 1-\frac{1}{2}\erf\left(\frac{\localmax(T)}{\sqrt{2}\sigma}\right) < e^{-3\mathcal{N}}, \label{stable_sigma}
\end{equation}
where we approximated the Higgs field strength distribution as a~Gaussian one \eqref{gauss_distribution}, thus $\erf$ is the Gauss error function i.e. integral of a~Gaussian distribution. The bound \eqref{stable_sigma} is given by the requirement that the fraction of the volume of the Universe occupied by the field strength of the Higgs field belonging the basin of the attraction $\UVbasin$ of the high energy minimum is lower then the inverse of the number of the Hubble horizons developed during inflation i.e. $e^{3\mathcal{N}}$ and correspond to the scenario in which no horizon with field strength of the Higgs field larger then the local maximum of the potential $\localmax$ is present in our past light cone. In the plot \ref{stable_sigma} we have used the value $\mathcal{N}=50$. Finally, the formation of Higgs domain walls is possible only for parameters satisfying the first condition $1.2 \localmax (T) > \sigma$, however not the one in equation \eqref{stable_sigma}.

Assuming that the inhomogeneous configuration of the Higgs field was formed during inflation we can use the expression \eqref{inflationary} in order to connect the standard deviation $\sigma$ with the value of the Hubble $H_I$ parameter during inflation. Described bounds expressed in therms of $H_I$ instead of $\sigma$ are presented in the figure \ref{thermal_Hmax}. 

\begin{figure}[!ht]
\subfloat[]{\label{thermal_RHplus}
\includegraphics[width=0.5 \textwidth]{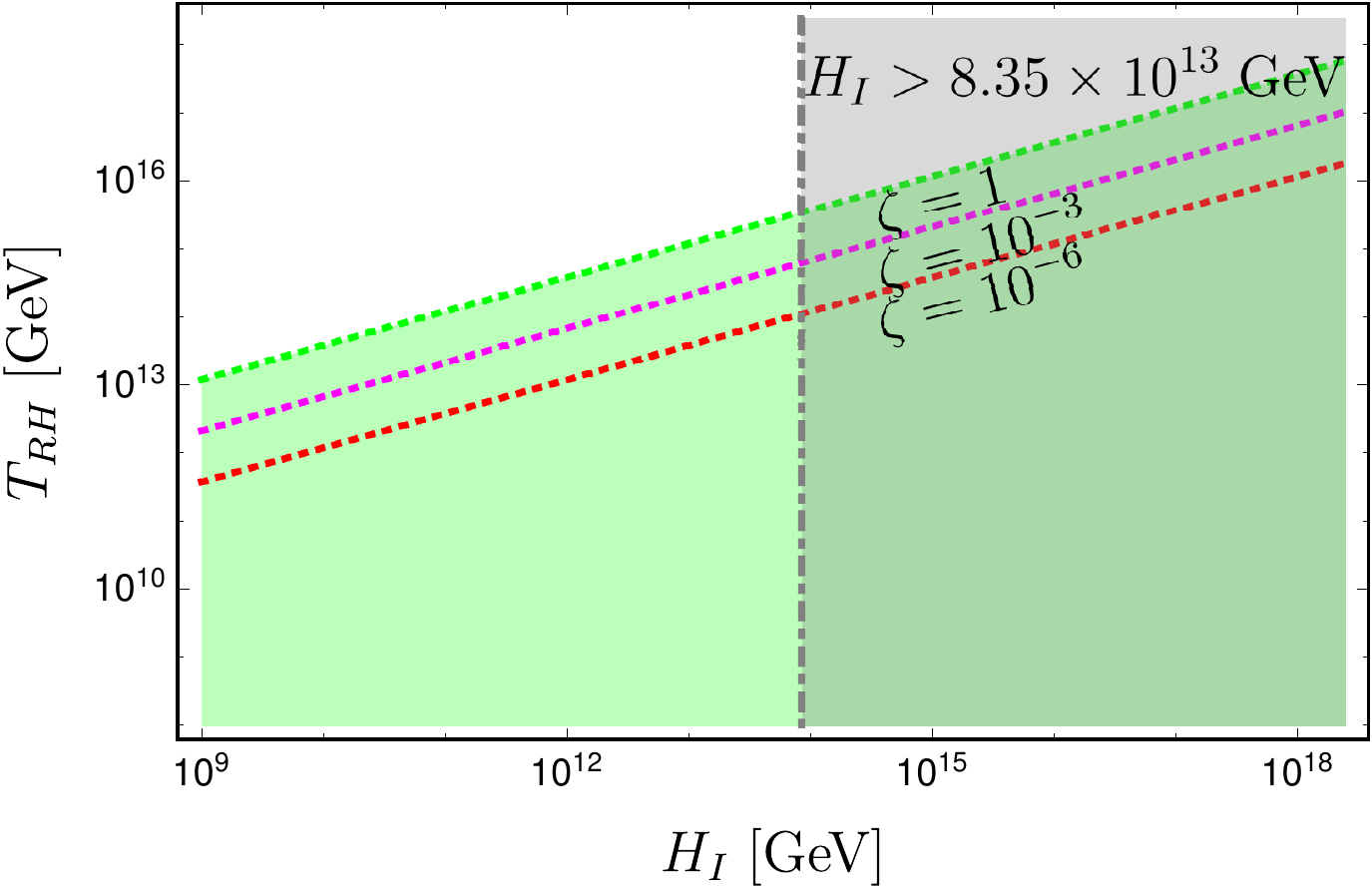}
 }
 \subfloat[]{\label{thermal_RHplusT}
\includegraphics[width=0.5 \textwidth]{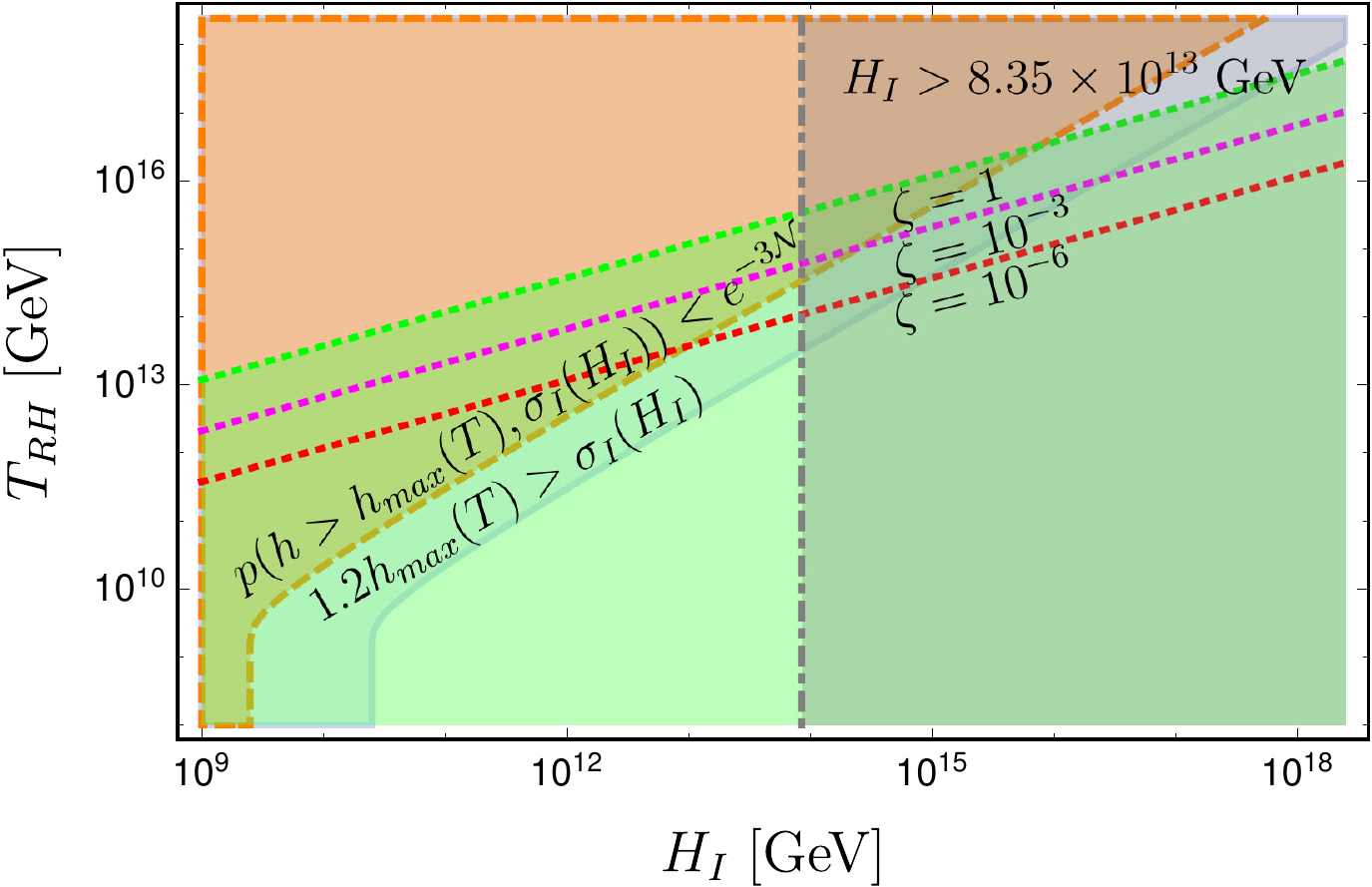}
 }
\caption{Range of reheating temperatures (green) for three values of the efficiency parameter $\zeta$: $\zeta=1$ (green line), $\zeta=10^{-3}$ (magenta line) and $\zeta=10^{-6}$ (red line) \protect\subref{thermal_RHplus}. Extrapolated region of the parameter space \ref{thermal_Hmax} for which the formation of Higgs domain walls is possible under assumptions: $\sigma_I \sim \frac{H_I}{2 \pi} \sqrt{\mathcal{N}}$ and $1.2 \localmax >\sigma_I$. 
\protect\label{thermal_max}}
\end{figure}

Most inflationary models predict that due to inflation the Universe become empty and cold, with the energy density dominated by the potential energy density of the inflaton field. To resume the standard cosmological evolution the energy density of the inflaton has to be transferred to other degrees of freedom. This transformation from the inflationary Universe to the radiation dominated Universe took place in the epoch called (p)reheating, during which 
interactions of the inflaton field produced particles %, especially those from the Standard Model enlarging
increasing the entropy of the Universe. Various inflationary models predict different scenarios of this process and different reheating temperatures $T_{RH}$ (defined as a temperature of the thermal bath of the Standard Model particles just after the the end of the reheating). Using energy conservation we can bound from above the reheating temperature with the Hubble parameter $H_I$ using the fact that the energy density at the end of inflation was ${\rho_{cr}}_I = 3 {M_{Pl}}^2 {H_I}^2$ had to be larger then the energy density of the thermal bath
\begin{equation}
T_{RH} < \left(\frac{30}{\pi^2 g_*} \right)^\frac{1}{4} \sqrt{M_{Pl} H_I}. %sprawdzić
\end{equation}
%Saturation of this bound is unlikely, because particles from beyond the Standard Model (at least dark matter) were produced and the energy density of the Standard Model particles was red-shifted by the expansion of the Universe during the reheating epoch.
We will parametrize our ignorance of the details of reheating by defining 
\begin{equation}
\zeta := \frac{\rho_{\text{SM}}}{{\rho_{cr}}_I}
\end{equation}
that simply describies its efficiency. Figure~\ref{thermal_RHplus} shows values of the reheating temperature $T_{RH}$ for efficiencies $\zeta=1$, $\zeta=10^{-3}$ and $\zeta=10^{-6}$ as a~function of the Hubble parameter during inflation $H_I$.

Comparison of these bounds with the estimated range of parameter space for which the formation of Higgs domain walls is possible (previously plotted in \ref{thermal_Hmax}) is presented in the figure \ref{thermal_RHplusT}. It can be deduced from the plot in the figure \ref{thermal_RHplusT} that the existence of domain walls decaying to the electroweak vacuum for $H_I > 10^{17}\; \textrm{GeV}$ is excluded if the standard deviation of the distribution $\sigma_I \sim \frac{H_I}{2 \pi} \sqrt{\mathcal{N}}$ is assumed. Moreover, the evolution for low temperatures is possible only if the efficiency of the reheating was low. In addition we have indicated as a~gray region on \ref{thermal_RHplusT} the present bound on the value of the Hubble parameter $H_I$ during single-field slow-roll inflation $H_I < 8,35 \times 10^{13}\; \textrm{GeV}$ coming from the scalar-to-tensor ratio bound $r<0.08$ from the Planck satellite measurement \cite{Ade:2015xua}. Basing on the plot \ref{thermal_RHplusT} one may naively presume that usually used in the literature \cite{Espinosa:2015qea,Espinosa:2016nld} bound \eqref{stable_sigma} on the value of the Hubble parameter during inflation $H_I$ may be weaken by the order of magnitude if the evolution of domain walls was taken under consideration. However, as we will see in the Section \ref{decay_time} this is not true, because in contrast to the previously investigated scenarios of Higgs domain walls in the vacuum background state, the dynamics of the Higgs field in the thermal background changes in time due to decreasing of the temperature of the background. The shape of the effective potential that drives the evolution of the network of Higgs domain walls evolve in time $t$ (especially the position of the local maximum $\localmax(T)$ becomes the time dependent quantity due to time dependence of the temperature $T(t)$). As a result, the only method that can reliably investigate mentioned hypotheses are lattice simulations which take into account the evolution of the temperature due to expansion of the Universe.

\section{Results of numerical simulations\label{decay_time}}
We have performed numerical simulations on the lattice of the size $256^3$ of the evolution of networks of Higgs domain walls in the background of the thermal state with temperatures ranging form $10^{10}\; \textrm{GeV}$ to $10^{17}\; \textrm{GeV}$. Under the assumption that no particles were produced during the simulated period of time, so the entropy was constant, one finds that the temperature scales as $T \propto a^{-1}$ with the expansion of the Universe, where $a(t)$ is the scale factor. As we have shown in Section \ref{wall}, the width of Higgs domain walls $w_{\text{SM}}(T)$ scales inversely proportional to the temperature $T$, so it stays nearly constant in the expending Universe. Due to this coincidence we were allowed to use the original eom instead of the modified equation proposed in the PRS algorithm \cite{Press:1989yh}. Basing on results presented in Section \ref{wall} we have chosen the lattice constant $l$ to be equal to the inverse of the initial temperature $T_{start}$ and the initial value of the scale factor $a_{start}=\frac{1}{4}$ which gives the width of the order of $w_{SM} = 40\, l$. Performing our simulation we have assumed that the evolution of Higgs domain walls took place in the radiation domination epoch. 

\begin{figure}[!ht]
\subfloat[]{\label{thermal_T=1e10}
\includegraphics[width=0.5 \textwidth]{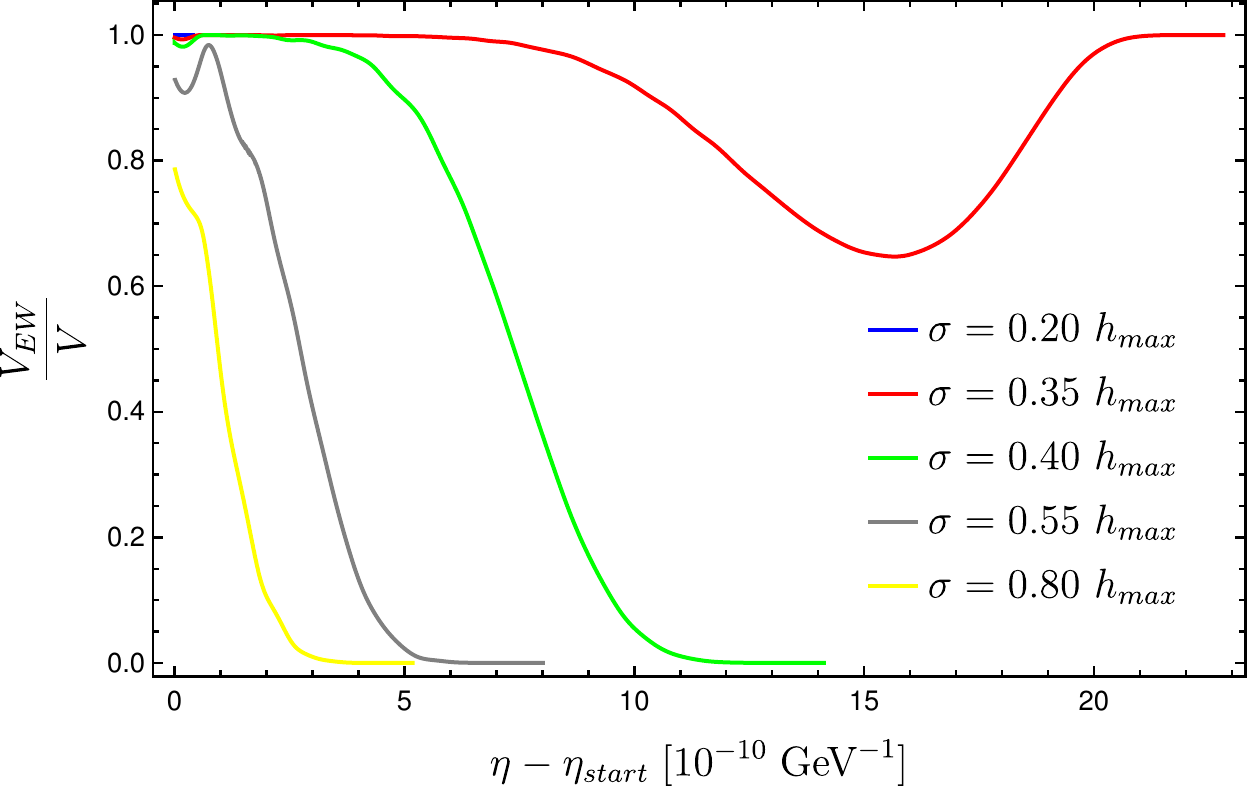}
 }
 \subfloat[]{\label{thermal_T=1e17}
\includegraphics[width=0.5 \textwidth]{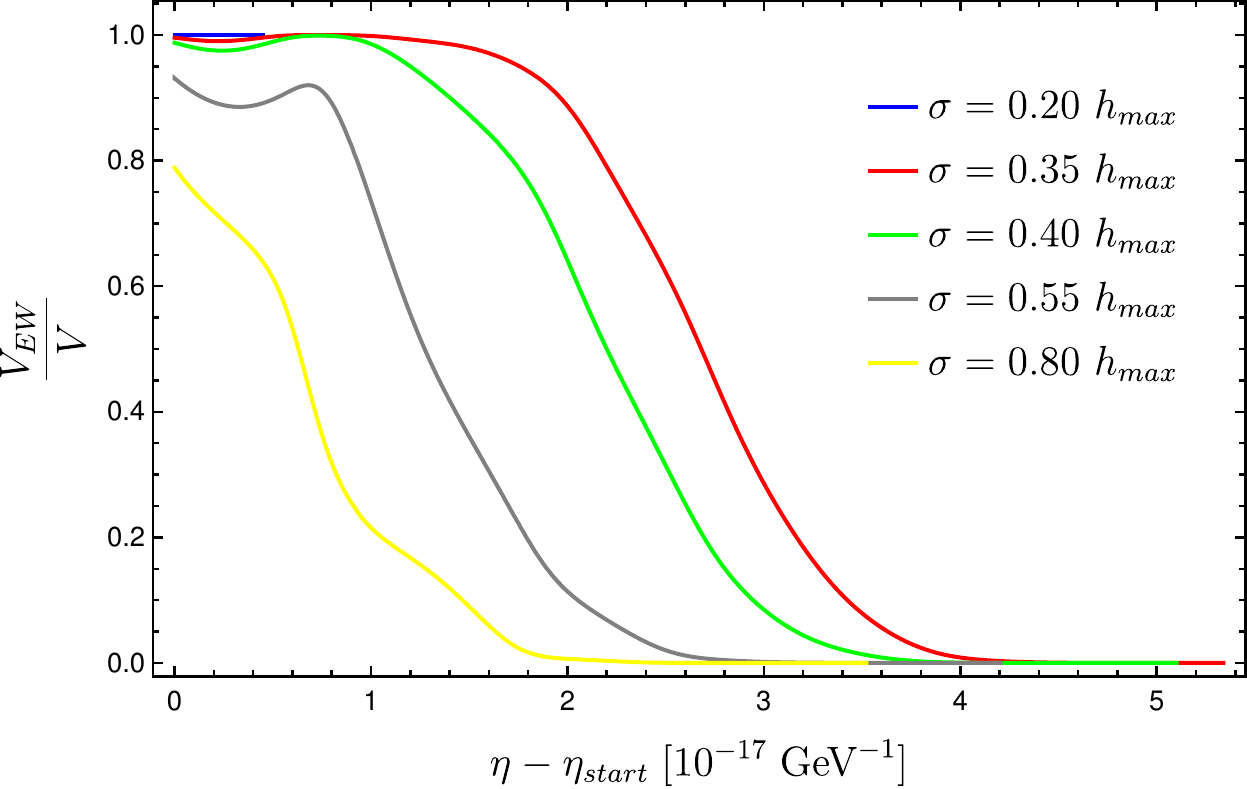}
 }
\caption{Time dependence of the fraction of lattice sites occupied by the field on the electroweak side of the barrier $\frac{V_{\text{EW}}}{V}$ for various values of the standard deviation $\sigma$ of the initial distribution for two values of initial temperature, respectively: $T_{start}=10^{10}\; \textrm{GeV}$ \protect\subref{thermal_T=1e10} and $T_{start}=10^{17}\; \textrm{GeV}$ \protect\subref{thermal_T=1e17}. \protect\label{thermal_results}}
\end{figure}

Results of our simulations are presented in the figure \ref{thermal_results}, where we have plotted the time dependence of the ratio $V_{\text{EW}}$ of the number of lattice points $V_{EW}$ with field strengths belonging to the basin of attraction of the electroweak vacuum $\EWbasin$ to the total number of lattice points $V$ for two values of the initial temperature equal to $T_{start}=10^{10}\; \textrm{GeV}$ and $T_{start}=10^{17}\; \textrm{GeV}$.

Our numerical studies show that the Higgs domain walls in the thermal background are significantly more unstable than in the zero temperature case
and their networks are more likely to decay into the experimentally excluded high field strength minimum $\minUV$. We can trace the source of this to two effects. Firstly, described in the section \ref{SM_potential} dependence of the position of the local maximum $\localmax(T)$ on the temperature $T$, so on the conformal time $\eta$, leads to dynamical decrease of the range of the basin of the attraction of the electroweak vacuum $\EWbasin$ during the evolution of domain walls. As a result, some patches of the Universe occupied by the field strength belonging to $\EWbasin$ at some early time may lay in the basin of attraction of the high field strength minimum  $\UVbasin$ at some latter time. If temperature $T$ of the thermal background drops down faster than the domain walls evolve, the Higgs field will finally land at the high field strength minimum $\minUV$ of the moving potential barrier. Secondly, due to damping (described in Section \ref{SM_potential}) of thermal corrections for field strengths $h$ larger then the value of the temperature $T$, the thermally corrected effective potential of the Standard Model is much steeper on the high field strength minimum side of the local maximum. As we pointer in our earlier papers, the asymmetry of the potential may have crucial role in the dynamics of cosmological domain walls. 

\begin{figure}[!ht]
\centering
\includegraphics[width=0.8 \textwidth]{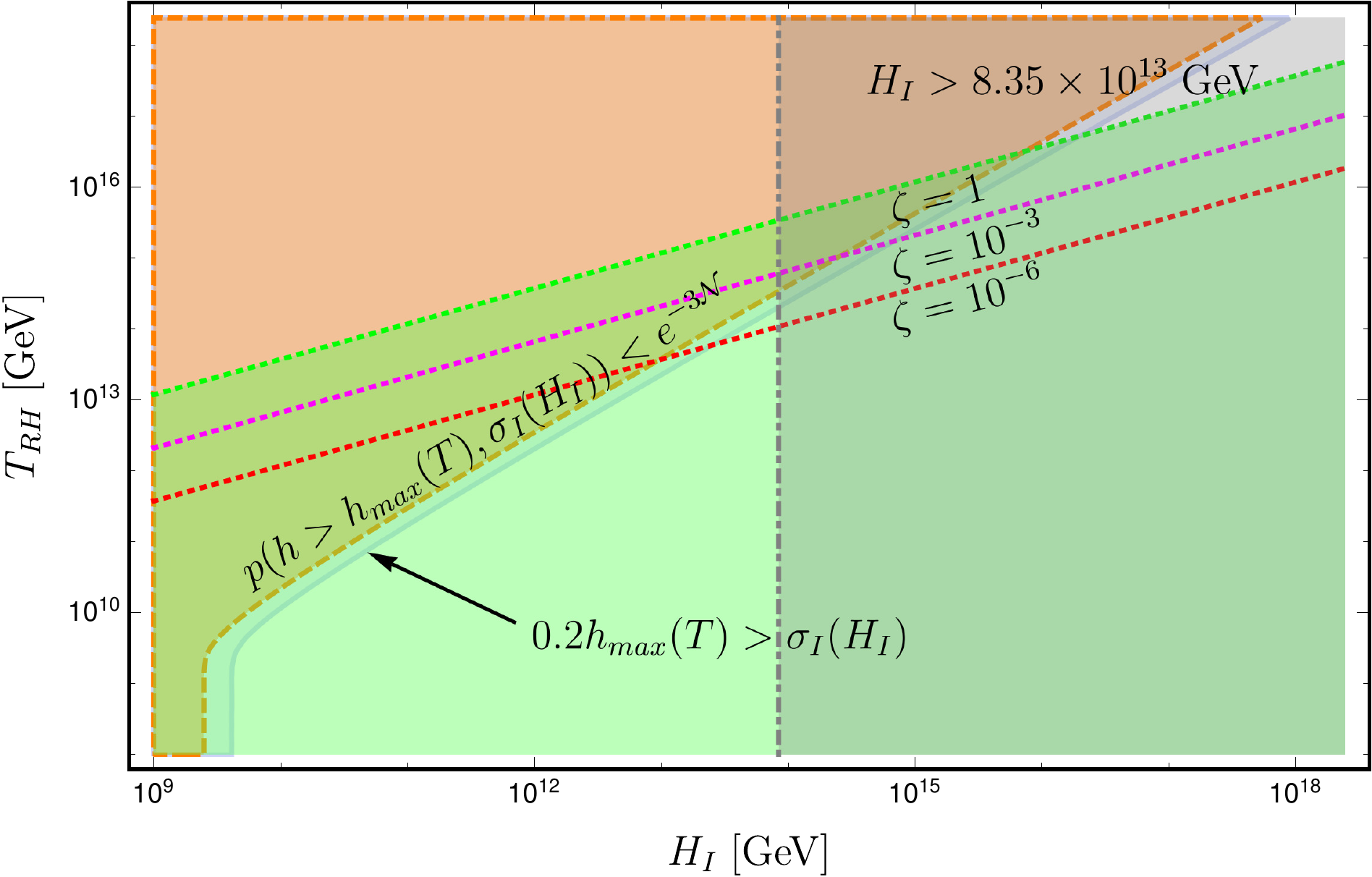}
\caption{Region of the parameter space \ref{thermal_Hmax} for which the formation of Higgs domain walls is possible under assumptions: $\sigma_I \sim \frac{H_I}{2 \pi} \sqrt{\mathcal{N}}$ and $0.2 \localmax >\sigma_I$. For points in the orange region the electroweak vacuum is chosen as the field is not pushed beyond the barrier by vacuum fluctuation. In case of $0.2\localmax>\sigma_I$ the field will end its evolution in the high field strength vacuum. The blue region represents part of the parameter space where the walls are produced but quickly decay leaving the field in the electrowek vacuum. The other lines represent reheating and are the same as in Fig \ref{thermal_RHplus}. \protect\label{thermal_RHplusTnew}}
\end{figure}

We summarise the results of our simulations in figure~\ref{thermal_RHplusTnew} (analogous to figure~\ref{thermal_RHplusT} in which we take into account the high instability of Higgs domain walls in the background of the thermal bath. Adopting $0.2 \localmax (T) > \sigma_I= \frac{H_I}{2 \pi} \sqrt{\mathcal{N}}$ one obtains the bound on the reheating temperature $T_{RH}$ and the Hubble parameter during inflation $H_I$ which leads to evolution of domain walls decaying to the electroweak vacuum $\minEW$.  The allowed parameter range is highly reduced comparing to the previous estimate from figure~\ref{thermal_RHplusT} which was based on the studies in the case of zero temperature background.

\section{Summary\label{summary}}
In this paper we investigate the influence of thermal corrections on the evolution of Higgs domain walls. Extrapolating our knowledge of the present Universe one finds that it has to be denser and hotter in the past. Thus, it is important to understand the effect of the background of the thermal state of the Standard Model particles on the dynamics of Higgs field. Our main result is that thermal corrections significantly change the dynamics of the Higgs field comparing with the vacuum case considered in our previous papers~\cite{Krajewski:2016vbr,Krajewski:2017czs}.

In this paper use Matsubara's imaginary time formalism in order to accommodate thermal corrections to the effective potential of the Standard Model. The shape of the potential is strongly modified at field strengths lower then the temperature of the thermal bath $T$. Especially, the position of the local maximum $\localmax$ separating the maxima increases proportionally to $T$ for temperatures higher then $10^{10}\; \textrm{GeV}$. On the other hand, contribution from thermal corrections is highly suppressed for higher field strengths and the neighbourhood of the global minimum is not modified. 

In section \ref{wall} we described properties of domain walls interpolating between the electroweak minimum of the thermally corrected effective potential and the second minimum located at high field strengths.
We determined the width of domain walls as a~function of the temperature $T$ of the thermal bath. We have found that for temperatures higher than $10^{10}\; \textrm{GeV}$ the width decreases inversely proportionality with $T$. This observation suggests that the thermal corrections shift the characteristic energy scale of the dynamics of Higgs domain walls.

Networks of cosmological domain walls could be produced in the early Universe by a~variety of processes. In this paper we concentrated on the possibility that Higgs field fluctuations over the potential barrier separating two minima of the potential were generated during cosmological inflation. Inflationary models predict configurations of the Higgs field well approximated by the Gauss distribution with the standard deviation $\sigma_I$ proportional to the Hubble constant during inflation $H_I$. The energy density of the early Universe after inflation was dominated by the potential energy of the inflaton field. In order to produced the present Universe the energy of the inflaton field had to be transferred to other degrees of freedom, especially Standard Model particles. This process is called reheating and its final state is the Universe filled with the thermal bath of the particles with temperature $T_{RH}$. The evolution of the Higgs field expectation value during inflation and (p)reheating has become an important issue after the existence of the high field strength global minimum in the Higgs potential was shown for central values of the measured parameters. In this work we are interested in the later evolution in the radiation domination era and the possible formation of topological defects---domain walls. In section \ref{initial} (assuming the inflationary scenario) we determine the parameter range of $H_I$ and $T_{RH}$ for which the formation of the Higgs domain walls was possible. Moreover, using naive extrapolation of results obtained in \cite{Krajewski:2016vbr,Krajewski:2017czs}, we estimate the bound on the $H_I$ from the condition of the proper final state of the evolution of the Higgs filed.

However, the situation in the presence of the thermal bath differs significantly from previously considered case of the background vacuum state. The shape of the effective potential of the Higgs field changes dynamically in time (especially the position of the local maximum $\localmax$ decreases) due to cooling down of the Universe with its expansion. Thus, the only reliable tool one can use to investigate the dynamics of the Higgs domain walls in this case is numerical lattice simulations. Section \ref{decay_time} presents results of numerical simulations based on the PRS algorithm which we have used previously in \cite{Krajewski:2016vbr,Krajewski:2017czs}. Simulations have shown that the Higgs domain walls in the background of the thermal state are much more unstable and decay in conformal time $\mathcal{O}(10) {T_0}^{-1}$ where $T_0$ is the initial value of the temperature. Moreover, we have found that much lower amplitude of fluctuations (the standard deviation of the initial distribution $\sigma_I$) is allowed by the requirement that the network should decay to electroweak minimum. Thus, Higgs domain walls had to be rare objects in the early Universe. As a result, Higgs domain walls decayed shortly after formation and constituted only a~tiny fraction of the energy density of the Universe, so they did not influenced the cosmological observables if they were formed at temperatures higher then $10^{10}\ \textrm{GeV}$.

Basing on results of our numerical simulations we can conclude that the naive bound on the Hubble parameter during inflation is not weakened significantly by taking into account of the dynamics of Higgs domain walls. Thus, one will obtain the valid estimate of the scale of parameters when studying inflationary models just by excluding the fluctuations of the Higgs field over the potential barrier after reaheating. 

\section*{Acknowledgments}
This work has been supported by the National Science Centre, Poland grants DEC-2012/04/A/ST2/00099, 2017/27/B/ST2/02531, 2014/13/N/ST2/02712 and 2016/23/N/ST2/03111 and MNiSW grant IP2015 043174 and the United Kingdom STFC Grant ST/P000258/1.

\bibliographystyle{elsarticle-num}
\bibliography{HDWTB}
\end{document}